\documentclass[12pt]{article}
\usepackage[pdftex]{graphicx} 
\usepackage{epsfig}
\usepackage{comment}
\usepackage{latexsym}
\usepackage{color}
\usepackage{amsfonts}
\usepackage{amsmath}
\usepackage{mathrsfs}

\newcommand{\mysquare}[0]{\raise-.2ex\hbox{{\Large$\Box$}}}
\def\lsim{\mathrel{\rlap {\raise.5ex\hbox{$ < $}}
{\lower.5ex\hbox{$\sim$}}}}
\def\gsim{\mathrel{\rlap {\raise.5ex\hbox{$ > $}}
{\lower.5ex\hbox{$\sim$}}}} \topmargin -1.5cm \textheight=22.5cm \textwidth=16.5cm
\catcode`\@=11
\newcount\hour
\newcount\minute
\newtoks\amorpm
\hour=\time\divide\hour by60 \minute=\time{\multiply\hour by60
\global\advance\minute by-\hour}
\edef\standardtime{{\ifnum\hour<12 \global\amorpm={am}%
        \else\global\amorpm={pm}\advance\hour by-12 \fi
        \ifnum\hour=0 \hour=12 \fi
        \number\hour:\ifnum\minute<10 0\fi\number\minute\the\amorpm}}
\edef\militarytime{\number\hour:\ifnum\minute<10 0\fi\number\minute}
\def\draftlabel#1{{\@bsphack\if@filesw {\let\thepage\relax
   \xdef\@gtempa{\write\@auxout{\string
      \newlabel{#1}{{\@currentlabel}{\thepage}}}}}\@gtempa
   \if@nobreak \ifvmode\nobreak\fi\fi\fi\@esphack}
        \gdef\@eqnlabel{#1}}
\def\@eqnlabel{}
\def\@vacuum{}
\def\draftmarginnote#1{\marginpar{\raggedright\scriptsize\tt#1}}
\def\draft{\oddsidemargin -.2truein
        \def\@oddfoot{\sl preliminary draft \hfil
        \rm\thepage\hfil\sl\today\quad\militarytime}
        \let\@evenfoot\@oddfoot \overfullrule 3pt
        \let\label=\draftlabel
        \let\marginnote=\draftmarginnote
   \def\@eqnnum{(\theequation)\rlap{\k

 ern\marginparsep\tt\@eqnlabel}%
\global\let\@eqnlabel\@vacuum}  }

\newcommand{\be}[0]{\begin{equation}}
\newcommand{\ee}[0]{\end{equation}}
\newcommand{\ba}[0]{\begin{eqnarray}}
\newcommand{\ea}[0]{\end{eqnarray}}

\def\bs{\begin{subequations}}
\def\es{\end{subequations}}

\def\thebibliography#1{%
\vskip 0.5cm \centerline{\bf \Large References}
\list{%
[\arabic{enumi}]}{\settowidth\labelwidth{[#1]} \leftmargin\labelwidth
\advance\leftmargin\labelsep
\usecounter{enumi}}
\def\newblock{\hskip .11em plus .33em minus .07em}
\sloppy\clubpenalty4000\widowpenalty4000 \sfcode`\.=1000\relax}

\renewcommand{\theequation}{\arabic{section}.\arabic{equation}}

\renewcommand{\section}{\setcounter{equation}{0}\@startsection
{section}{1}{0mm}{-\baselineskip}{0.5\baselineskip} {\normalfont\Large\bfseries}}

\renewcommand{\subsection}{\@startsection
{subsection}{2}{0mm}{-\baselineskip}{0.5\baselineskip} {\normalfont\large\bfseries}}

\renewcommand{\subsubsection}{\@startsection
{subsubsection}{3}{0mm}{-\baselineskip}{0.5\baselineskip}
{\normalfont\normalsize\slshape}}

\usepackage{amssymb}
\usepackage{amsthm}
\usepackage{amsmath}
\usepackage{amssymb,amsfonts}
\usepackage{graphicx}
\usepackage{cite}

\renewcommand{\and}{\mbox{and}}

\begin{document}
\begin{titlepage}
\begin{flushright}
CERN-PH-TH/2015-038 
\end{flushright}

\vspace{2cm}

\begin{centering}
{\bf\Huge Universality of radiative corrections to gauge couplings for strings with spontaneously broken supersymmetry 
}
\\

\vspace{1.5cm}

 { \LARGE Ioannis Florakis$^\ast$}

\vspace{1cm}

{\large Theory Division, CERN\\ \vspace{1mm}
  1211 Geneva 23, Switzerland\\ 
}

 \vspace{2cm}

{\bf\Large Abstract}\\
\vspace{0.5cm}

\begin{flushleft}
I review recent work on computing radiative corrections to non-abelian gauge couplings in four-dimensional heterotic vacua with spontaneously broken supersymmetry. The prototype models can be considered as K3 surfaces with additional Scherk-Schwarz fluxes inducing the spontaneous $\mathcal{N}=2 \to \mathcal{N}=0$ breaking. Remarkably, although the gauge thresholds are no longer BPS protected and receive contributions also from the excitations of the RNS sector, their difference is still exactly computable and universal. 
\end{flushleft}

\vskip 0.2cm
\emph{Based on a talk presented at the DISCRETE 2014 conference at King's College London}

\end{centering}

\begin{quote}

\noindent 
\end{quote}
\vspace{5pt} \vfill  \vskip.1mm{\small \small \small \noindent \ e-mail:\ \ \texttt{ioannis.florakis@cern.ch}\\
}

\end{titlepage}
\newpage
\setcounter{footnote}{0}
\renewcommand{\thefootnote}{\arabic{footnote}}
 \setlength{\baselineskip}{.7cm} \setlength{\parskip}{.2cm}

\setcounter{section}{0}


\section{Introduction}

The last twenty years have been marked by a significant progress in String Theory and, consequently, in String Phenomenology, with the construction of several semi-realistic vacua incorporating the salient features of the MSSM. In particular, the low energy effective action with $\mathcal N=1$ supersymmetry has been successfully reconstructed at tree level. However, an eventual quantitative comparison with low energy data clearly necessitates the incorporation of quantum corrections to various couplings and, therefore, the incorporation of loop corrections in both the string length $\ell_s$ and the string coupling $g_s$ are at the centre of attention in the modern string literature.

In fact, gravitational and gauge coupling corrections at the one loop level in $g_s$ have been extensively studied in the literature, for vacua with at least one unbroken supersymmetry. Until very recently, however, the study of quantum corrections to gauge couplings in vacua with broken supersymmetry still constituted an unexplored terrain. One of the main difficulties  one faces whenever supersymmetry is broken in string theory is the appearance of tachyonic excitations, that destabilise the classical vacuum and invalidate the perturbative treatment. This is possible in a theory with an infinite and exponentially growing number of states such as String Theory and, in the case of spontaneously broken supersymmetry, it  is  closely related with the Hagedorn problem of String Thermodynamics \cite{Atick:1988si}. Several proposals for stabilising the winding tachyon have appeared in the literature \cite{Angelantonj:2006ut,Angelantonj:2008fz,Florakis:2010ty}, involving setups with specialised fluxes, orbifolds and orientifolds, but they will not be discussed here. Rather, I will assume that one is working in a region of moduli space where supersymmetry is spontaneously broken but the classical vacuum is stable. In this case, it is meaningful and important to study one-loop\footnote{The problem of incorporating higher-loop corrections is more involved, due to the necessity of incorporating the backreaction of the one-loop tadpole on the classical vacuum.} radiative corrections to couplings in the low energy effective action.

In this note, I will review the first calculation \cite{AFT} of one-loop corrections to non-abelian gauge couplings in four dimensional heterotic vacua with spontaneously broken supersymmetry and discuss a remarkable and highly non-trivial universality property of differences of such thresholds. The prototype model that I will consider is the four-dimensional heterotic vacuum with $\mathcal N=2$ supersymmetry obtained by compactifying the ten-dimensional ${\rm E}_8\times {\rm E}_8$ heterotic string on ${\rm K}3\times T^2$. In the following, I will assume a trivial Wilson line background so that the non-abelian gauge group be enhanced to ${\rm E}_8\times {\rm E}_7$. The $\mathcal{N}=2$ supersymmetry is then spontaneously broken to  $\mathcal{N}=0$ by turning on a Scherk-Schwarz flux \cite{Scherk:1978ta}, in a way that admits an exactly tractable worldsheet CFT description in terms of a freely-acting $\mathbb{Z}_2$ orbifold \cite{Rohm:1983aq,Kounnas:1988ye,Ferrara:1988jx,Kounnas:1989dk}. The resulting model is a four-dimensional heterotic vacuum with ${\rm SO}(16)\times{\rm SO}(12)$ non-abelian gauge group factor and spontaneously broken supersymmetry $\mathcal{N}=2\to\mathcal{N}=0$.

The reason for studying this particular prototype model may be appreciated if one realises that, even in more realistic vacua with spontaneously broken $\mathcal N=1$ supersymmetry, the only dependence of threshold corrections on the compactification moduli is contained in $\mathcal{N}=2\to \mathcal{N}=0$ subsectors\footnote{ Clearly, the contribution of pure $\mathcal N=1$ subsectors does not involve the Narain lattice and is, hence, moduli independent. Its explicit value depends on the infrared renormalisation scheme and will not be discussed here. Is is, however, important to stress that it may be computed using the recently developed techniques of \cite{AFP1,AFP2,AFP3}. For short reviews, see \cite{Florakis:2013ura,Pioline:2014bra}.}. Consequently, the resulting expression for the gauge thresholds obtained using this model is  immediately applicable to more realistic setups, since their $\mathcal N=2$ subsectors are precisely of the above prototype form. In particular, they should apply to the models of \cite{Faraggi:2014eoa}, recently proposed in the context of addressing the decompactification problem. It is hoped that the recent activity in the analysis of non-supersymmetric string vacua \cite{Blaszczyk:2014qoa,Abel:2015oxa}, an integral part of which necessarily involves the study of quantum corrections to gauge couplings,  may lead to a better understanding of string theoretical implications for low energy phenomenology.


\section{The prototype model}

One may construct a four-dimensional $\mathcal N=2$ heterotic vacuum on ${\rm K}3\times T^2$ by working in the singular limit of K3, realised as a $T^4/\mathbb Z_N$ orbifold with $N=2,3,4,6$. Denoting $Z^1, Z^2$ the complexified coordinates of $T^4$, the crystallographic rotation of the orbifold action reads
\begin{equation}
	\begin{split}
	& Z^1 \rightarrow e^{2\pi i/N}\,Z^1 \,,\\
	& Z^2 \rightarrow e^{-2\pi i/N}\,Z^2 \,.\\
	\end{split} 
\end{equation}
The $\mathcal N=2$ supersymmetry is then spontaneously broken by turning on a Scherk-Schwarz flux, realised as a freely-acting $\mathbb Z_2'$ orbifold with element
\begin{equation}
	v' = (-1)^{F_{\rm s.t.}+F_1+F_2}\,\delta \,,
	\label{SSaction}
\end{equation}
where $F_{\rm s.t.}$ is the spacetime fermion number, $F_1$, $F_2$ are the `fermion numbers' associated to the two original ${\rm E}_8$'s, and $\delta$ is an order-two shift along a cycle of the untwisted $T^2$. The restriction to a trivial Wilson line background ensures that the classical vacuum is stable, even though supersymmetry is spontaneously broken, and the theory enjoys a non-abelian ${\rm SO}(16)\times{\rm SO}(12)$ gauge symmetry.

The modular covariant one-loop partition function of the theory reads
\begin{equation}
\begin{split}
\mathcal{Z}=&\tfrac{1}{2} \sum_{H,G=0}^1\tfrac{1}{N} \sum_{h,g=0}^{N-1}\left[\tfrac{1}{2}\sum_{a,b=0}^1 (-)^{a+b}
\vartheta \left[^{a/2}_{b/2}\right]^2 \,
\vartheta \left[^{a/2+h/N}_{b/2+g/N} \right]\,
\vartheta \left[^{a/2-h/N}_{b/2-g/N}\right]
\right]
\\
&\qquad\times \left[\tfrac{1}{2}\sum_{k,\ell=0}^1
\bar\vartheta \left[^{k/2}_{\ell/2}\right]^6\,
\bar\vartheta \left[^{k/2+h/N}_{\ell/2+g/N} \right]\,
\bar\vartheta \left[^{k/2-h/N}_{\ell/2-g/N} \right]
\right]\,
\left[\tfrac{1}{2}\sum_{r,s=0}^1\bar\vartheta \left[^{r/2}_{s/2} \right]^8\right] 
\\
&\qquad\times \,\frac{1}{\eta^{12}\bar\eta^{24}}
\,(-)^{H(b+\ell+s)+G(a+k+r)+HG}\, \varGamma_{2,2} \left[^H_G \right] 
\,\varLambda^{\rm K3} \left[^h_g\right] \,,
\end{split}\label{partFunction}
\end{equation}
and is expressed in terms of the genus-one Jacobi theta and Dedekind functions.
A few comments are in order, regarding the notation. The first line in \eqref{partFunction} contains the contribution of the left-moving RNS sector, whereas the second line is associated to the gauge bundle. Finally, the third line involves the oscillator contributions $\eta^{-12}\,\bar\eta^{-24}$, the phase realising the Scherk-Schwarz action \eqref{SSaction} and lattice factors associated to the compactification. In particular, $h,H$ label the various (un)twisted sectors of the $\mathbb Z_N$ and $\mathbb Z_2'$ orbifolds, respectively. Similarly, the summation over $g,G$ imposes the orbifold projections, in each case. 

$\varGamma_{2,2}[^H_G]$ is the shifted Narain lattice associated to the untwisted $T^2$ given by
\begin{equation}
	\varGamma_{2,2}[^H_G] = \tau_2 \sum_{m_i, n^i \in \mathbb Z} e^{i\pi G m_1}\, q^{\frac{1}{4}|P_L|^2} \,\bar{q}^{\frac{1}{4}|P_R|^2} \,,
\end{equation}
corresponding to a momentum shift along the first $T^2$ cycle. The left- and right- moving lattice momenta are given by
\begin{equation}
	P_L = \frac{m_2-U m_1+\bar T(\hat{n}^1+U \hat{n}^2)}{\sqrt{T_2\,U_2}} \ ,\qquad P_R = \frac{m_2-U m_1+  T(\hat{n}^1+U \hat{n}^2)}{\sqrt{T_2\,U_2}} \,,
\end{equation}
where $\hat{n}^i = n^i +\delta_1^i\,H/2$, and $T,U$ are the K\"ahler and complex structure moduli of $T^2$, respectively. The implementation of the Scherk-Schwarz mechanism responsible for the spontaneous supersymmetry breaking in terms of the $\mathbb Z_2'$ orbifold is achieved by the insertion of the phase $(-)^{H(b+\ell+s)+G(a+k+r)+HG}$ together with the above momentum shift on the $\varGamma_{2,2}[^H_G]$ lattice. This can be most conveniently seen by noticing that $(-)^{a+k+r}$ is precisely identified with the operator $(-)^{F_{\rm s.t.}+F_1+F_2}$ since $a$ is the spacetime fermion number, and $k,r$ are the `fermion' numbers associated to the ${\rm SO}(16)$ spinor representations associated to the original ${\rm E}_8$'s. The additional part $(-)^{H(b+\ell+s)+HG}$ is straightforwardly identified with its modular completion. On the other hand, the momentum shift in $\varGamma_{2,2}[^H_G]$  precisely corresponds to a left-right symmetric shift in the first $T^2$ coordinate $X^1\to X^1+\pi R$.  Coupling the operator of fermion numbers  to the $\varGamma_{2,2}$ lattice with a momentum shift, hence, precisely generates the $\mathbb Z_2'$ orbifold action \eqref{SSaction}.

Finally, the twisted $(4,4)$ lattice partition function is given by
\begin{equation}
\varLambda^{\rm K3} \left[^h_g \right] = \left\{
\begin{array}{cl}
\varGamma_{4,4} &\qquad {\rm for}\ (h,g)=(0,0)\,,
\\[10pt]
 \frac{ k \left[^h_g \right]\, |\eta |^{12}}{ \left| \vartheta \left[ {1/2+h/N \atop 1/2 +g/N}\right]\, 
\vartheta \left[ {1/2-h/N \atop 1/2 - g/N}\right] \right|^2} &\qquad {\rm for}\ (h,g)\not= (0,0)\,,
\end{array}
\right.
\end{equation}
with $\varGamma_{4,4}$ being the conventional Narain lattice associated to the $T^4$,  $k \left[^0_g \right] = 16\, \sin^4 (\pi g/N)$ counting the number of $\mathbb Z_N$ fixed points, and the remaining $k \left[^h_g \right]$'s with $h\neq 0$ are determined by modular invariance.  

The spectrum of the theory may be straightforwardly derived by direct inspection of the partition function \eqref{partFunction}. In particular, the freely-acting nature of the $\mathbb Z_2'$ orbifold implies that no state is projected out but, rather, states carrying non-trivial charge under $F_{\rm s.t.}+F_1+F_2$ acquire non-trivial masses. In particular, the two gravitini of the $\mathcal N=2$ theory are degenerate and their mass is given by
\begin{equation}
	m_{3/2}^2 = \frac{|U|^2}{T_2 U_2} \,.
\end{equation}

A very important difference to the supersymmetric case, particular to the spontaneous breaking of supersymmetry, and originating from the modification of the GSO projection by the Scherk-Schwarz boost, is the presence of charged BPS states that may become massless at special loci in the bulk of the Narain moduli space. Indeed, consider first the general formula for the left- and right- moving masses of string states
\begin{equation}
	\begin{split}
		m_L^2 = |P_L|^2 + 4 N_{\rm  osc}-2 \,,\\
		m_R^2 = |P_R|^2 + 4 \bar N_{\rm osc}-4 \,,
	\end{split}\label{massFormulae}
\end{equation} 
where $N_{\rm osc}, \bar N_{\rm osc}$ are the left- and right- moving oscillator excitations and the constants $-2$, $-4$ are associated to the worldsheet vacuum energies of the RNS and bosonic (right-moving) sectors, respectively. In the supersymmetric case, the mass of BPS states is dictated by the $\mathcal N=2$ central charge, $m_{{\rm BPS}}^2=|P_L|^2$, and correspond to the tower of Kaluza-Klein momentum and winding excitations of the RNS ground state $N_{\rm osc}=1/2$. Together with level-matching, this implies that states charged under the ${\rm E}_8\times {\rm E}_7$ gauge group may only become massless if $\bar N_{\rm osc}=1$, implying that they lie in the BPS subsector $m_i n^i =0$. The latter constraint may be solved explicitly and it is straightforward to show that  extra charged massless states can only occur in the boundary of the $T,U$ moduli space.

On the other hand, when supersymmetry is spontaneously broken by the Scherk-Schwarz mechanism, the mass of BPS states is no longer given by $|P_L|^2$ and the modification of the GSO projection allows for the presence of charged states with $N_{\rm osc}=0$, $\bar N_{\rm osc}=1$, charged in the bi-fundamental of ${\rm SO}(16)\times{\rm SO}(12)$. The string mass formul\ae\,  \eqref{massFormulae} then imply that the BPS mass is modified to $m_{\rm BPS}^2=|P_R|^2$ and, together with level matching, show that these states lie in the BPS subsector $m_i \hat{n}^i = -1/2$. Hence, these charged BPS states carry non-trivial momentum and winding charges around the cycle of $T^2$ where the momentum shift acts, $m_1=-2\hat{n}^1=\pm 1$ and $m_2=n^2=0$, and become massless at loci in the bulk of the Narain moduli space lying in the T-duality orbit generated by $T=2U$. Such BPS states indeed exist in the perturbative string spectrum of the model and may be seen to arise from the $a=h=0$, $H=1$ sector. Their origin in the partition function \eqref{partFunction} is most clearly expressed in terms of ${\rm SO}(2n)$ characters
\begin{equation}
	O_4 O_4 \bar V_{12} \bar O_4 \bar V_{16}\times \tfrac{1}{2} \left( \varGamma_{2,2}[^1_0]+\varGamma_{2,2}[^1_1] \right) \,,
	\label{extraMassless}
\end{equation}
and their mass is explicitly given by
\begin{equation}
	m_{\rm BPS}^2 = \frac{|T/2-U|^2}{T_2 U_2} = |P_R|^2 \,.
\end{equation}
The fact that the mass of these charged states is no longer dictated by the $\mathcal N=2$ central charge $P_L$ is a result of the mass deformation induced by the Scherk-Schwarz flux.


\section{$\mathcal N=2$ Gauged supergravity}

In this section, I will discuss the Scherk-Schwarz mechanism inducing the spontaneous breaking of supersymmetry in the prototype model of the previous section from the low energy supergravity perspective. 

In Field Theory, the Scherk-Schwarz mechanism constitutes a deformation of the fields of the theory by a symmetry operator $Q$, such that fields $\varPhi(X_5)$ depending on a compact coordinate $X_5$ of radius $R$ acquire a non-trivial monodromy as one encircles the compact direction $\varPhi(X_5+2\pi R) = e^{2\pi iQ}\,\varPhi(X_5)$. Consequently, this non-periodicity induces a shift in the Fourier frequencies and the Kaluza-Klein spectrum of charged states is modified $m\to m+Q$. In particular, for a charged massless scalar $\varPhi$, the ground state in the Kaluza-Klein spectrum is now shifted to $m_{\rm KK} = |Q|/R$.

In String Theory, the Scherk-Schwarz mechanism acts in a very similar way, as a deformation of the vertex operators by the symmetry operator $Q$. It can be interpreted as a constant field strength background $F_{IJ}$ for the circle ${\rm U}(1)$ taking values in the internal manifold and corresponds to a deformation of the worldsheet  action by
\begin{equation}
	\delta S_{\rm 2d} = \int d^2 z\, F_{IJ} (\psi^I\,\psi^J -X^I  \overset{\leftrightarrow}{\partial} X^J) \, \bar\partial X_5 \,.
	\label{SSdeform}
\end{equation}
It is clear that the operator $\psi^I \psi^J- X^I \overset{\leftrightarrow}{\partial} X^J$ is simply the rotation operator in the internal $I$-$J$ plane, whereas $\bar\partial X_5$ is the translation operator along the compact $X_5$ direction. Their coupling precisely induces the Scherk-Schwarz boost on the lorentzian charge lattice. Although the above deformation of the worldsheet Lagrangian is not quadratic in the fields, it is nevertheless integrable for special quantised values of $F_{IJ}$. Indeed, the Lagrangian deformation explicitly involves the bosonic field $X^I$ without its derivative, which is not a well-defined conformal field. This reflects the fact that the operator $X^I \overset{\leftrightarrow}{\partial} X^J$ does not have a well-defined action on the compact toroidal internal space for a generic continuous rotation angles $F_{IJ}$. However, the deformation does become integrable for special quantised values of $F_{IJ}$, corresponding to crystallographic symmetries of the torus. This is precisely the worldsheet manifestation of flux quantisation. In fact, for these quantised values of the flux, the deformation \eqref{SSdeform} may be shown to be absorbed into the free kinetic terms of the worldsheet action, by an appropriate redefinition of the boundary conditions of the worldsheet fields (for a recent discussion, see \cite{Condeescu:2012sp,Condeescu:2013yma}).

The Scherk-Schwarz deformation of the theory corresponds to a flat gauging of $\mathcal N=2$ supergravity and the effective action up to the two-derivative level is completely fixed by the couplings among vector- and hyper- multiplets. For simplicity, I will consider the $T^4/\mathbb Z_2$ realisation of the K3 surface so that the scalar manifold may be obtained by a simple $\mathbb Z_2$ truncation of the $\mathcal N=4$ supergravity one, and has the form
\begin{equation}
	\left(\frac{{\rm SU}(1,1)}{{\rm U}(1)}\right)_S \times \left(\frac{{\rm SO}(2,2)}{{\rm SO}(2)\times{\rm SO}(2)}\right)_{T,U} \times \left(\frac{{\rm SO}(4,4+n)}{{\rm SO}(4)\times{\rm SO}(4+n)}\right) \,.
\end{equation}
where the first coset corresponds to the axio-dilaton scalar $S$, the second coset is parametrised by the $T,U$ moduli of the $T^2$, and the last coset is a quaternionic manifold of hypermultiplets containing the K3 moduli together with the infinite number ($n=\infty$) of BPS multiplets of the theory.


In order to analyse the effective action of the non-supersymmetric, non-tachyonic model \eqref{partFunction} and the effect of extra massless states at special points in the classical $T,U$ moduli space one retains only the low lying charged BPS states that may become massless. For simplicity, the K3 moduli  will be frozen at their minima and an additional $\mathbb Z_2$ truncation will be performed in order to express the theory in $\mathcal N=1$ language\footnote{This is consistent since the two gravitini of the gauged $\mathcal N=2$ theory are degenerate in mass and the extra charged massless states arise in the untwisted $\mathcal N=1$ subsector, hence, surviving the truncation.}. The scalar manifold  then becomes
\begin{equation}
	\left(\frac{{\rm SU}(1,1)}{{\rm U}(1)}\right)_S \times \left(\frac{{\rm SO}(2,2)}{{\rm SO}(2)\times{\rm SO}(2)}\right)_{T,U} \times \left(\frac{{\rm SO}(2,n_+)}{{\rm SO}(2)\times{\rm SO}(n_+)}\right)_{Z_+^A} \times \left(\frac{{\rm SO}(2,n_-)}{{\rm SO}(2)\times{\rm SO}(n_-)}\right)_{Z_-^A} \,.
\end{equation}
Let us now identify the charged BPS states of interest, at the level of the string spectrum. From the point of view of the truncated $\mathcal N=1$ theory, they arise from the sector
\begin{equation}
	O_2\,O_2\,O_2\,O_2\ \bar V_{10}\,\bar O_2\,\bar O_2\,\bar O_2 \,\bar V_{16}\times \tfrac{1}{2}\left( \varGamma_{2,2}[^1_0]+\varGamma_{2,2}[^1_1] \right) \,.
\end{equation}
The relevant BPS states may then be denoted as $Z^A_{\pm}$, where $A=(a,\hat a)$ is an index in the bi-fundamental $({\bf 10},{\bf 16})$ of the ${\rm SO}(10)\times{\rm SO}(16)$ gauge group. The subscript $\pm$ splits the states according to their momentum and winding charges
\begin{equation}
	\begin{split}
	Z^A_{+} \ &:\qquad m_1=2n_1= +1\ , \qquad m_2=n_2=0 \,,\\
	Z^A_{-} \ &:\qquad m_1=2n_1= -1\ , \qquad m_2=n_2=0 \,.
	\end{split}
\end{equation}
The structure constants $f_{ABC}$ of the gauging may be straightforwardly computed at the $\mathcal N=2$ string level by simple three-point correlation functions, $f_{ABC}=\langle \mathcal{V}_A \mathcal{V}_B \mathcal{V}_C\rangle$.  The ${\rm SU}(1,1)/{\rm U}(1)$ and ${\rm SO}(2,n)/{\rm SO}(2)\times{\rm SO}(n)$ coset conditions may be explicitly solved and, upon matching the $\mathcal N=1$ gravitino mass term $m_{3/2}$ given in terms of the structure constants of the $\mathcal N=2$ gauging with $e^{K/2}W$, it is straightforward to identify the K\"ahler potential
\begin{equation}
	K=-\log S_2\, T_2\, U_2 \left[1-2|\vec Z_+|^2+(\vec Z_+^2)(\vec Z_+^\ast)^2\right]\left[1-2|\vec Z_-|^2+(\vec Z_-^2)(\vec Z_-^\ast)^2\right] \,,
\end{equation}
and superpotential of the $\mathcal N=1$ truncation
\begin{equation}
	W = \sqrt{2}\left[-2(T+2U)\,\vec Z_+\cdot \vec Z_-+2\,U(1+\vec Z_+^2)(1+\vec Z_-^2)\right]\,,
\end{equation}
in terms of the physical scalar fields $S,T,U,Z^A_\pm$. The scalar potential is then obtained from
\begin{equation}
	V= e^K\left[(K^{-1})^{i\bar\jmath}(W_i+K_i W)(\bar W_{\bar\jmath}+K_{\bar\jmath}\bar W)-3|W|^2\right]\,.
\end{equation}
The point $Z_{\pm}^A=0$ is an extremum of the potential, with $V|_{Z_\pm^A=0}=0$. The gravitino mass around this point is found to be
\begin{equation}
	m_{3/2}^2 =e^{K}|W|^2= \frac{|U|^2}{S_2 T_2 U_2}\,,
\end{equation}
which correctly reproduces the perturbative string spectrum formula. On the other hand, the masses of the $Z^A_\pm$ fields can be determined after appropriate diagonalisation and one finds
\begin{equation}
		\begin{split}
	\chi_1^A \ &:\qquad M_1^2 = \frac{1}{S_2}\,\frac{|T/2- U|^2}{T_2 U_2}=|P_R|^2=|P_L|^2-2 \ , \qquad \qquad \qquad   \     m_1=2n_1=\pm 1\,,\\
	\chi_2^A \ &:\qquad M_2^2 = \frac{1}{S_2}\,\left[\frac{|T/2-U|^2}{T_2 U_2}+4\right]=|P_R|^2+4=|P_L|^2+2 \ , \ \  \quad  \quad m_1=2n_1=\pm 1\,,\\
	\chi_3^A \ &:\qquad M_3^2 = \frac{1}{S_2}\,\left[\frac{|T/2+ \bar U|^2}{T_2 U_2}+2\right]=|P_R|^2=|P_L|^2+2 \ , \qquad \ \ \ \ \ \ \, m_1=-2n_1=\pm 1\,,\\
	\chi_4^A \ &:\qquad M_4^2 = \frac{1}{S_2}\,\left[\frac{|T/2-3 U|^2}{T_2 U_2}+4\right]=|P_R|^2+4=|P_L|^2-2 \ , \qquad m_1=6n_1=\pm 3\,,
	\end{split}
\end{equation}
where
\begin{equation}
	\begin{split}
	&\chi_1^A =\tfrac{1}{2}\,\textrm{Im}(Z_+^A - Z_-^A) \ , \qquad \chi_2^A=\tfrac{1}{2}\,\textrm{Im}(Z_+^A + Z_-^A) \,,\\
	&\chi_3^A=\tfrac{1}{2}\,\textrm{Re}(Z_+^A - Z_-^A) \ ,\qquad \chi_4^A=\tfrac{1}{2}\,\textrm{Re}(Z_+^A + Z_-^A) \,.
	\end{split}
\end{equation}
By inspection of the above mass spectrum, it is clear that only the states $\chi_1^A$ can become massless. We therefore retain only their contribution ($\chi_1\neq 0$) and freeze all other fields to their minima $\chi_2^A=\chi_3^A=\chi_4^A=0$, impling the identification $Z_+^A = i \chi^A$, $Z_-^A = -i\chi^A$. The scalar potential is then found to be
\begin{equation}
	V=\frac{1}{S_2}\,\frac{\vec\chi^2}{(1-\vec\chi^2)^4}\left[ \frac{|T-2\bar U|^2}{T_2 U_2}\left(1+(\vec\chi^2)^2\right)+2\,\frac{|T+2U|^2}{T_2 U_2}\,\vec\chi^2\right] \,.
\end{equation}
This expression may be further simplified by performing the analytic field redefinition
\begin{equation}
	\Phi^A = \frac{\chi^A}{1-\chi^B\chi^B}\,,
\end{equation}
upon which the scalar potential is cast into the form
\begin{equation}
	V=\frac{1}{S_2}\,\left[\frac{|T-2\bar U|^2}{T_2 U_2}\,\vec\Phi^2+2\left(\frac{|T-2\bar U|^2+|T/2+U|^2}{T_2 U_2}\right)\,(\vec\Phi^2)^2\right]\,.
	\label{scalarPot}
\end{equation}
From the structure of the scalar potential \eqref{scalarPot}, it is straightforward to see that the theory possesses a stable minimum at $\Phi^A=0$, where the potential vanishes. This reflects the `no-scale' structure of the Scherk-Schwarz gauging, which implies that $S,T,U$ remain moduli at tree level and may be stabilised eventually through higher loop corrections. Notice that the no-scale structure of the potential is consistent with the fact that the worldsheet CFT is exactly solvable at the minimum, with the $T,U$ moduli corresponding to marginal deformations of the current-current type entering the shifted Narain lattice $\varGamma_{2,2}$. The presence of extra charged  massless states naturally introduces logarithmic infrared singularities in the structure of gauge threshold corrections, as will be discussed in the next section.


\section{Gauge threshold corrections and universality}

Having discussed the prototype model and the supergravity description of extra charged massless states, I will briefly review the universality arising in differences of gauge couplings in the case of unbroken supersymmetry and then focus on the appearance of a remarkable and unexpected universality in the case where supersymmetry is spontaneously broken by the Scherk-Schwarz flux \eqref{SSaction}. 

The running of the gauge coupling  associated to a gauge factor $\mathcal G$ at one loop takes the form
\begin{equation}
	\frac{16\pi^2}{g_{\mathcal G}^2(\mu)} = \frac{16\pi^2}{g_s^2}+\beta_{\mathcal G}\,\log\frac{M_s^2}{\mu^2}+\Delta_{\mathcal G} \,,
\end{equation}
where the second term in the r.h.s. is proportional to the beta function coefficient $\beta_{\mathcal G}$ and corresponds to the field theory result. The third term $\Delta_{\mathcal G}$ is the correction due to the infinite tower of massive string states running in the loop and is known as the threshold correction. 

The one loop correction to the gauge coupling is computed in string theory by considering the two point CFT correlator of the vertex operators of gauge bosons
\begin{equation}
	\mathcal V^a(z,\bar z) = A_\mu^a (\partial X^\mu+ip\cdot\psi \psi^\mu)\,\tilde J^a(\bar z)\,e^{ip\cdot X}\,,
\end{equation}
with $\tilde J^a(\bar z)$ being the Kac-Moody currents realising the loop algebra
\begin{equation}
	\tilde J^a(\bar z)\,\tilde J^b(0) = k\frac{\delta^{ab}}{\bar z^2} +i{f^{ab}}_c \frac{\tilde J^c(0)}{\bar z}+{\rm reg}\,,
\end{equation}
where ${f^{ab}}_c$ are the structure constants of the Lie algebra of $\mathcal G$. After integrating the position $(z,\bar z)$ over the worldsheet torus, one is left  to perform the modular integral over the moduli space of gauge inequivalent metrics, parametrised by the complex structure parameter $\tau$, over a fundamental domain $\mathcal F$
\begin{equation}
	\frac{16\pi^2}{g^2_{\mathcal G}} = {\rm RN} \int_{\mathcal F} \frac{d^2\tau}{\tau_2^2} \int_{\rm  torus} d^2 z\, \langle \mathcal V^a(z,\bar z)\,\mathcal V^a(0)\rangle_{\rm CFT}\,,
\end{equation}
and RN stands for a renormalisation prescription for treating the infrared divergences due to the presence of massless states. We shall henceforth suppress the explicit display of the `RN' symbol in subsequent integrals, assuming the appropriate renormalisation prescription of \cite{AFP1,AFP2, AFP3}. Schematically, the threshold takes the form
\begin{equation}
	\frac{16\pi^2}{g^2_{\mathcal G}} = \int_{\mathcal F} \frac{d^2\tau}{\tau_2^2} \sum_{\rm states} {\rm Str}\left(\frac{1}{12}-s^2\right)\left(Q^2_{\mathcal G}-\frac{1}{4\pi\tau_2}\right)\, q^{\frac{1}{4}|P_L|^2+N_{\rm osc}-\frac{1}{2}}\,\bar q^{\frac{1}{4}|P_R|^2+\bar N_{\rm osc}-1}\,,
	\label{generalStructure}
\end{equation}
where $s$ is the helicity operator, $Q_{\mathcal G}$ is a Cartan charge in the gauge group $\mathcal G$ and $q=e^{2\pi i\tau}$ is the nome. 

A few comments are in order about the structure of the above expression. The exponents of $q$ and $\bar q$ are simply the left- and right- moving conformal weights of the states running in the loop, respectively. In the rectangular region $\tau_2>1$ of the fundamental domain, the $\tau_1\in(-\frac{1}{2},\frac{1}{2})$ integration simply imposes level matching whereas $\tau_2$ plays the role of the field-theoretic Schwinger parameter. In the non-rectangular region $\tau_2<1$ of $\mathcal F$, the integral receives contribution also from non-level matched states, as demanded by modular invariance and unitarity. The supertrace operator ${\rm Str}(\frac{1}{12}-s^2)Q_\mathcal{G}^2$ is identified in the massless sector  with the field-theoretic beta function coefficient $\beta_{\mathcal G}$ and is associated to the contribution of the charged states running in the loop. The charge-independent term proportional to $1/\tau_2$, on the other hand, is associated to a non-1PI diagram due to the universal coupling of the dilaton, and arises as a result of a modular regularisation of short-distance divergences on the string worldsheet.

\subsection{Supersymmetric universality}

In the case when supersymmetry is unbroken, the $F^2$ term is BPS saturated and only BPS states contribute to the threshold. Using the fact that in this subsector $N_{\rm osc}=1/2$, it is straightforward to see from \eqref{generalStructure} that the left moving oscillators cancel out and one obtains a simple expression of the form
\begin{equation}
	\Delta_{\mathcal G}= \int_{\mathcal F} \frac{d^2\tau}{\tau_2^2} \,\varGamma_{2,2}(T,U)\,\varPhi(\bar\tau)\,,
	\label{simplicity}
\end{equation}
where the Narain lattice $\varGamma_{2,2}$ encodes the lattice sum over the left- and right- moving momenta and the remaining right-moving oscillator contributions, weighted with appropriate super- and group- trace coefficients, are encoded in the weakly almost holomorphic modular function $\varPhi$. The term `weakly holomorphic' refers to a simple pole in $\bar q$ arising from the bosonic vacuum $\bar N_{\rm osc}=0$ of the heterotic string. On the other hand, the term `almost holomorphic' refers to the breaking of holomorphy due to the explicit appearance of the $1/\tau_2$ term.

I will not discuss here the explicit evaluation of this modular integral, nor the explicit form of $\varPhi$ which is dependent on the choice of gauge group factor $\mathcal G$. Rather, I will consider the difference of gauge thresholds associated to two different gauge group factors $\mathcal G_1$ and $\mathcal G_2$. Since the universal dilaton diagram is independent of the choice of gauge group, the $1/\tau_2$ terms cancel in the difference, and one arrives at
\begin{equation}
	\Delta_{\mathcal G_1}-\Delta_{\mathcal G_2} =\int_{\mathcal F} \frac{d^2\tau}{\tau_2^2} \, \varGamma_{2,2}(T,U)\,C(\bar \tau)\,,
\end{equation}  
where $C(\bar\tau)$ is now a weakly holomorphic modular function. Its generic Fourier expansion reads
\begin{equation}
	C(\bar\tau) = \frac{c_{-1}}{\bar q}+c_0+c_1\,\bar q +\ldots
\end{equation}
Invoking a well-known theorem from Number Theory, stating that any weakly holomorphic modular form of non-positive weight is uniquely determined by the principal part of its $q$-expansion, we are immediately lead to write 
\begin{equation}
	C(\bar\tau) = c_{-1} \bar j(\bar\tau)+ c_0 \,,
\end{equation}
where $j(\tau)$ is the Hauptmodul of the modular group ${\rm SL}(2;\mathbb Z)$ known as the Klein $j$-invariant\footnote{The theorem guarantees that $j(\tau)$ is the unique modular invariant function with a simple pole in $q$ and I will conventionally define it with vanishing constant term in its Fourier expansion, {\emph i.e.} $j(\tau)=\frac{1}{q}+196884\,q+21493760\,q^2+\ldots$. Note that the coefficients of all positive powers in $q$ are uniquely determined by modularity and from the knowledge of the coefficient of the simple pole.}. It is now possible to fix both $c_{-1}$ and $c_0$ by a simple inspection of \eqref{generalStructure}. Notice first, that $c_{-1}$ corresponds to the bosonic right-moving ground-state of the heterotic string $\bar N_{\rm osc}=0$ which is always uncharged. Since the dilaton diagram cancelled out in the difference, only the difference of charges contributes in the difference of thresholds and, hence, $c_{-1}=0$. The constant, on the other hand is the contribution of the massless states $\bar N_{\rm osc}=1$ and is, hence, given by the difference of beta function coefficients for the two gauge groups, $C(\bar\tau)=\beta_{\mathcal G_1}-\beta_{\mathcal G_2}$. The resulting integral for the difference of thresholds then involves the Narain lattice alone and was computed in the seminal paper \cite{DKL}, yielding
\begin{equation}
	\Delta_{\mathcal G_1}-\Delta_{\mathcal G_2} = -(\beta_{\mathcal G_1}-\beta_{\mathcal G_2})\,\log\left( T_2 U_2\, |\eta(T)\,\eta(U)|^4\right) + {\rm const}.
	\label{supersymUniversality}
\end{equation}
This result is universal\footnote{Although certain exceptions do arise, for instance, when one deals with compactifications on non-factorisable tori, see \cite{Mayr:1993mq,Kiritsis:1998en}.} and, modulo the model dependent beta function prefactor, it is independent of the details of the string vacuum. 

It should be stressed that the universality structure crucially relied on the presence of unbroken supersymmetry which, for the moduli dependent contributions under consideration arising from the BPS subsector, the left-moving oscillators cancelled out and the integrand function $\varPhi$ was constrained to be almost holomorphic.

\subsection{Non-supersymmetric universality}

Now we consider the case when the Scherk-Schwarz flux is turned on and supersymmetry is spontaneously broken. In this case, the $F^2$ term receives contributions from all string states, notably including the non-BPS ones, and the previous simple expression  \eqref{simplicity} is no longer true. Rather, one encounters a sum of contributions ranging over the $\mathbb Z_2'$ orbifold sectors of the form
\begin{equation}
	\Delta_{\mathcal G} = \tfrac{1}{2}\int_{\mathcal F} \frac{d^2\tau}{\tau_2^2} \,\sum_{H,G=0}^{1} \varGamma_{2,2}[^H_G]\,\varPhi_{\mathcal G}[^H_G](\tau,\bar\tau) \,,
\end{equation}
where each orbifold sector $(H,G)$ is accompanied by a manifestly non-holomorphic, gauge factor-dependent function $\varPhi_{\mathcal G}[^H_G](\tau,\bar\tau)$ which is only modular with respect to the Hecke congruence subgroup $\varGamma_0(2)$ of the modular group.

To illustrate this point, one may display explicitly the modular integral of  the threshold for the ${\rm SO}(16)$ gauge group factor, with K3 realised as a $T^4/\mathbb Z_2$ orbifold. Explicitly, one finds \cite{AFT}
\begin{equation}
	\begin{split}
		&\Delta_{\rm SO(16)} = 2\int_{\mathcal F} \frac{d^2\tau}{\tau_2^2} \biggr\{ -\tfrac{1}{48}\,\varGamma_{2,2}[^0_0]\,\frac{\hat{\bar E}_2\,\bar{E}_4\,\bar{E}_6-\bar{E}_6^2}{\bar\eta^{24}}  \\
		&+\varGamma_{2,2}[^0_1] \biggr[-\tfrac{1}{1152}\,\frac{\varLambda^{\rm K3}[^0_0]}{\eta^{12}\bar\eta^{24}}\,(\vartheta_3^8-\vartheta_4^8)\,\bar\vartheta_3^4\,\bar\vartheta_4^4\,\left( (\hat{\bar{E}}_2-\bar\vartheta_3^4)\,\bar\vartheta_3^4\,\bar\vartheta_4^4+8\bar\eta^{12}\right)\biggr] \\
		&+\varGamma_{2,2}[^0_1]\biggr[-\tfrac{1}{96}\,\frac{\bar\vartheta_3^4\,\bar\vartheta_4^4\,(\bar\vartheta_3^4+\bar\vartheta_4^4)\,\left[(\hat{\bar{E}}_2-\bar\vartheta_3^4)\,\bar\vartheta_3^4\,\bar\vartheta_4^4+8\bar\eta^{12}\right]}{\bar\eta^{24}} \\
		&-\tfrac{1}{144}\,\frac{\vartheta_2^4(\vartheta_3^8-\vartheta_4^8)}{\eta^{12}}\,\frac{(\hat{\bar{E}}_2-\bar\vartheta_3^4)\,\bar\vartheta_3^4\,\bar\vartheta_4^4+8\bar\eta^{12}}{\bar\eta^{12}}\biggr]+ (S\cdot\tau)+(ST\cdot \tau)\biggr\} \,.
	\end{split}
	\label{complexity}
\end{equation}
The expression in the first line is separately invariant under ${\rm SL}(2;\mathbb Z)$ and corresponds to the BPS subsector, which effectively `feels' the unbroken $\mathcal N=2$ supersymmetry. The second line, on the other hand, is manifestly non-holomorphic and contains the dependence on the hypermultiplet moduli of K3.  The third line is again a BPS contribution arising in the $T^4/\mathbb{Z}_2$ realisation of K3, and is due to an `overlap' between the $\mathbb Z_2$ and $\mathbb Z_2'$ orbifolds rendering this subsector effectively supersymmetric. The last line represents a gauge factor-dependent, manifestly non-holomorphic contribution plus its images under $S$ and $ST$ transformations of ${\rm SL}(2;\mathbb Z)$.

I will not be discussing in this note the explicit evaluation of  the above modular integral, which produces a result in the form on an asymptotic expansion valid in the large volume region of moduli space. Instead, I will directly focus on the difference of thresholds by paralleling the supersymmetric discussion. \emph{A priori}, there is no reason to expect that the universality structure present in the supersymmetric case will persist also here, given that the simple property of (almost) holomorphy of  $\varPhi(\bar\tau)$ in \eqref{simplicity} is no longer present. Moreover, the explicit dependence of \eqref{complexity} on the K3 moduli is another reflection of the fact that the amplitude is no longer topological, but depends on the details of the K3 compactification.

Nevertheless, there is a number of simplifications that occur in the difference of thresholds for the ${\rm SO}(16)$ and ${\rm SO}(12)$ gauge factors. As in the supersymmetric case, the dilaton exchange diagrams cancel since they are independent of the choice of gauge group, and the $1/\tau_2$ contributions encoded in the almost holomorphic Eisenstein series $\hat{\bar E}_2$ cancel. Furthermore, the $h=g=0$ sector contribution proportional to $\varLambda^{\rm K3}[^0_0]$ is identical 
 for both gauge group factors and also cancels in the difference, eliminating the dependence on the hypermultiplet moduli. In addition, one may partially unfold \cite{AFP3} the fundamental domain $\mathcal F$ of ${\rm SL}(2;\mathbb Z)$ to the fundamental domain $\mathcal F_0(2)=\cup\{1,S,ST\}\cdot \mathcal F$ of the Hecke congruence subgroup $\varGamma_0(2)$, at the benefit of integrating a simpler expression - since its images under $S$ and $ST$ have been precisely absorbed into enlarging the fundamental domain. Schematically, one is left with
 \begin{equation}
 	\begin{split}
 	\Delta_{\rm SO(16)}-\Delta_{\rm SO(12)} &= ({\rm const.})\times \int_{\mathcal F}\frac{d^2\tau}{\tau_2^2}\,\varGamma_{2,2}[^0_0] \\
			&+\int_{\mathcal F_0(2)}\frac{d^2\tau}{\tau_2^2}\,\varGamma_{2,2}[^0_1]\,\sum_i\,({\rm holom.})_i\times({\rm anti-holom.})_i \,,
	\end{split}
 \end{equation}
where the first line on the r.h.s. corresponds to the BPS subsector contribution, whereas the second line arises from the manifestly non-holomorphic contributions and is a sum of terms involving products of holomorphic times anti-holomorphic contributions. In the case at hand, is has the explicit form
\begin{equation}
	\int_{\mathcal F_0(2)} \frac{d^2\tau}{\tau_2^2}\,\varGamma_{2,2}[^0_1]\left\{ -\frac{\vartheta_2^8\,|\vartheta_3^4+\vartheta_4^4|^2\,\bar\vartheta_3^4\bar\vartheta_4^4}{\eta^{12}\,\bar\eta^{24}}-\frac{\vartheta_2^4 \vartheta_3^4\,|\vartheta_2^4-\vartheta_4^4|^2\,\bar\vartheta_3^4\bar\vartheta_4^4}{\eta^{12}\,\bar\eta^{24}}+\frac{\vartheta_2^4 \vartheta_3^4\,|\vartheta_2^4+\vartheta_3^4|^2\,\bar\vartheta_3^4\bar\vartheta_4^4}{\eta^{12}\,\bar\eta^{24}}\right\} \,.
\end{equation}
Remarkably, this apparently non-holomorphic expression is reduced into a simple holomorphic one as a result of non-trivial reduced MSDS identities \cite{MSDS} and allows the result for the integral to be resummed into simple closed form. Indeed, by expanding this expression into ${\rm SO}(8)$ characters, one first notices that it factorises as $12(O_8^2\,V_8+3V_8^3)(\bar O_8^2 \bar V_8-\bar V_8^3)$. The non-trivial MSDS identity $\bar O_8^2 \bar V_8-\bar V_8^3=8$ then guarantees that the non-holomorphic contributions cancel against each other and the expression actually reduces to a purely holomorphic one, as a consequence of  MSDS spectral flow \cite{MSDS,Faraggi:2011aw}. The resulting $\varGamma_0(2)$ integral then simply becomes similar to the BPS-saturated ones
\begin{equation}
	\frac{1}{3}\int_{\mathcal F_0(2)}\frac{d^2\tau}{\tau_2^2} \,\varGamma_{2,2}[^0_1]\,\left(8-\frac{\vartheta_2^{12}}{\eta^{12}}\right) \,,
\end{equation}
and was computed in closed form in \cite{AFP4} where the underlying generalised Borcherds product formula for the $\varGamma_0(2)$ subgroup was derived. There are two important differences with the supersymmetric case. First, aside from the shifted Narain lattice factor, the integrand is now a holomorphic modular function of $\varGamma_0(2)$ rather than an anti-holomorphic one. This curious property implies that the difference of non-supersymmetric thresholds actually receives contributions from the bosonic (right moving) ground state and involves an infinite tower of oscillator excitations from the RNS sector, in a sense, being the left-right `mirror' of  BPS saturated amplitudes. The second difference lies in the fact that the holomorphic modular form of $\varGamma_0(2)$ is no longer a constant as in the supersymmetric case, but actually contains a $q$-pole at the cusp $\tau=0$. This is inherently related to the existence of the extra charged massless states \eqref{extraMassless} arising from the $N_{\rm osc}=0$, $\bar N_{\rm osc}=1$ sector.

Using the explicit results for the modular integrals discovered in \cite{AFP3,AFP4}, we find the following \emph{universal} result \cite{AFT}
\begin{equation}
	\begin{split}
	\Delta_{\rm SO(16)}-\Delta_{\rm SO(12)} = \alpha \,\log\,\left[ T_2 U_2\,|\eta(T)\,\eta(U)|^4\right] + \beta\,\log\left[T_2 U_2\,|\vartheta_4(T)\,\vartheta_2(U)|^4\right] \\
				+\gamma\,\log\bigr| j_2(T/2)- j_2(U)\bigr|^4 \,,
	\end{split}
	\label{universalResult}
\end{equation}
where $j_2(\tau)$ is the Hauptmodul of $\varGamma_0(2)$ and $\hat j_2(\tau)$ is its Atkin-Lehner transform. They can both be expressed in terms of the  genus-one Jacobi theta and Dedekind eta functions via
\begin{equation}
	j_2(\tau) = \left(\frac{\eta(\tau)}{\eta(2\tau)}\right)^{24}+24 \ ,\qquad \hat j_2(\tau) = j_2(-\tfrac{1}{2\tau}) = \left(\frac{\vartheta_2(\tau)}{\eta(\tau)}\right)^{12}+24 \,.
\end{equation}
Notice that the dependence on the details of the vacuum is only encoded in the coefficients $\alpha, \beta, \gamma$. It is a remarkable and highly non-trivial result, that the difference of thresholds can be cast into this simple closed-form universal expression involving known elliptic functions, irrespectively of the orbifold realisation of the ${\rm K}3\sim T^4/\mathbb{Z}_N$ surface.  In particular one finds \cite{AFT} that the triplet of coefficients $(\alpha,\beta,\gamma)$  is equal to $(72,-\frac{8}{3},\frac{2}{3})$ for $N=2,3$, equal to $\frac{5}{8}(72,-\frac{8}{3},\frac{16}{15})$ for $N=4$ and equal to $\frac{35}{144}(72,-\frac{8}{3},\frac{2}{3})$ for $N=6$. 

One then arrives at a manifestation of a universality structure for the difference of gauge thresholds in heterotic vacua where supersymmetry is spontaneously broken via  Scherk-Schwarz flux. Contrary to the supersymmetric case, this universality emerges only in differences of gauge thresholds is a remnant of the topological K3 universality present in the supersymmetric case and crucially relies on the spontaneous nature of the breaking. 

The effect of the Scherk-Schwarz mechanism is to deform the mass spectrum, breaking the perturbative ${\rm SO}(2,2)$ T-duality group down to a $\varGamma^0(2)_T\times\varGamma_0(2)_U$ subgroup. Indeed, the first two terms on the r.h.s. of \eqref{universalResult} precisely reflect the latter breaking of the T-duality group and are to be considered as freely-acting deformations of \eqref{supersymUniversality}. Similar terms appear in cases of partial spontaneous supersymmetry breaking $\mathcal N=4\to\mathcal N=2$. The third term, however, is particular to the spontaneous supersymmetry breaking down to $\mathcal N=0$ and can be identified to the presence of the extra charged massless states \eqref{extraMassless}. Indeed, it develops logarithmic singularities at precisely the loci $T=2U$ and its $\varGamma_0(2)$ images, where the latter charged BPS states become massless. This observation has an important consequence in providing a physical interpretation of the coefficients $\alpha,\beta,\gamma$. Namely, whereas $\alpha$ and $\beta$ are again interpreted as differences in beta function coefficients, $\gamma$ is proportional to the jump in the those beta function coefficients, due to the presence of additional massless states at $T=2U$.


\section{The one-loop effective potential at large volume}

Before ending the discussion, it is instructive to make a comment on the form of the vacuum energy of the prototype model, namely, the one-loop contribution to the effective potential.  Indeed, the breaking of supersymmetry implies that the modular integral of the partition function is no longer vanishing. On the other hand, the spontaneous nature of the breaking drastically restricts its volume dependence. The integrand is manifestly non-holomorphic, and one may at best obtain a large volume expansion of the result. Unfolding the fundamental domain $\mathcal F_0(2)$ against the Narain lattice proceeds in a straightforward fashion after decomposing the shifted Narain lattice into $\mathcal F_0(2)$ orbits. In particular, the vanishing orbit corresponding to $H=G=0$ vanishes identically as $\mathcal N=2$ supersymmetry is effectively recovered and this eliminates a dependence of the potential on positive powers of the volume of $T^2$. Dropping exponentially suppressed terms, it is straightforward to see that the result arises entirely from the degenerate orbit projected down to its zero modes and has the form
\begin{equation}
	V_{\rm eff} = -\frac{1}{T_2}\left[2N_{Q=1}\,E^\star(3;U)+2^3\left(\tfrac{1}{2}\,N_{Q=0}-N_{Q=1}\right)\,E^\star(3;2U)\right] +\mathcal O(e^{-|\delta|\sqrt{T_2}})\,,
\end{equation}
where $N_Q$ counts the number of massless bosons minus the number of massless fermions with charge $Q=(F_{\rm s.t.}+F_1+F_2)\,{\rm mod}\,2$, and
\begin{equation}
	E^\star(s;U)=\frac{\varGamma(s)}{2\pi^s} \sum_{(m,n)\neq(0,0)}\frac{U_2^s}{|m+Un|^{2s}}\,,
\end{equation}
is a properly normalised non-holomorphic Eisenstein series of weight 0. The spontaneous nature of the supersymmetry breaking, hence, implies that the effective potential is cast in the form
\begin{equation}
	V_{\rm eff} = -\beta(U)\,m_{3/2}^4+ \mathcal O(e^{-|\xi| m_{3/2}}) \,,
\end{equation}
in terms of the supersymmetry breaking scale $m_{3/2}$. This structure is compatible with the absence of $M_s^2 m_{3/2}^2$ terms from the effective potential, required  for the determination via radiative corrections of the no-scale modulus to lie around the TeV scale \cite{Ferrara:1994kg}.


 \section*{Acknowledgements}
 I would like to thank my collaborators C. Angelantonj and M. Tsulaia for a very enjoyable collaboration in which the results reviewed here were obtained. It is also my pleasure to thank A. Faraggi and N. Mavromatos for their kind invitation to present this work at DISCRETE 2014.


\end{document}